\documentclass[12pt]{iopart}
\usepackage[dvips]{graphicx}

\usepackage{hyperref}

\usepackage{color}
\definecolor{contentlistcolor}{rgb}{.5,.5,.7}

\def\mathvecfont#1{\textbf{\em #1}}
\newcommand{\myvec}[1]{\mathvecfont{#1}}

\newcommand{\ci}{\mathrm{i}}

\begin{document}
\title[Revivals of quantum wave-packets in graphene]{Revivals of quantum wave-packets in graphene}
\author{Viktor Krueckl$^1$ and Tobias Kramer$^{1,2}$}
\address{$^1$ Institute for Theoretical Physics, University of Regensburg, 93040 Regensburg, Germany}
\address{$^2$ Department of Physics, Harvard University, Cambridge, MA~02138, USA}
\ead{viktor.krueckl@physik.uni-r.de\\tobias.kramer@physik.uni-r.de}

\begin{abstract}
We investigate the propagation of wave-packets on graphene in a perpendicular magnetic field and
the appearance of collapses and revivals in the time-evolution of an initially localised wave-packet.
The wave-packet evolution in graphene differs drastically from the one in an electron gas and shows
a rich revival structure similar to the dynamics of highly excited Rydberg states.
We present a novel numerical wave-packet propagation scheme in order to solve the effective single-particle 
Dirac-Hamiltonian of graphene and show how the collapse and revival dynamics is affected by the presence of disorder.
Our effective numerical method is of general interest for the solution of the Dirac equation in the presence of potentials
and magnetic fields.
\end{abstract}

%73.23.-b Electronic transport in mesoscopic systems
%81.05.Uw Carbon, diamond, graphite 
%03.65.Sq Semiclassical theories and applications 
%42.50.Md Optical transient phenomena: quantum beats, photon echo, free-induction decay, dephasings and revivals, optical nutation, and self-induced transparency 

\submitto{\NJP}

\contentsline {section}{\numberline {1}Introduction}{2}{section.1}
\contentsline {section}{\numberline {2}Wave-packet propagation for the graphene-system}{2}{section.2}
\contentsline {section}{\numberline {3}Wave-packet revivals}{5}{section.3}
\contentsline {subsection}{\numberline {3.1}Model and observables}{5}{subsection.3.1}
\contentsline {subsection}{\numberline {3.2}Revivals in the centre-of-mass position}{7}{subsection.3.2}
\contentsline {subsubsection}{\numberline {3.2.1}Collapse}{9}{figure.4}
\contentsline {subsubsection}{\numberline {3.2.2}Revivals}{10}{subsubsection.3.2.2}
\contentsline {subsection}{\numberline {3.3}Revivals in the autocorrelation function}{10}{subsection.3.3}
\contentsline {subsection}{\numberline {3.4}Fractional revivals}{12}{subsection.3.4}
\contentsline {subsection}{\numberline {3.5}Effect of impurities}{14}{subsection.3.5}
\contentsline {section}{\numberline {4}Conclusion}{17}{section.4}
\contentsline {section}{\numberline {A}Classical Propagation}{17}{section.A}

\maketitle

\section{Introduction}

Since its experimental discovery, graphene has become a hot topic in solid state physics~\cite{Novoselov10222004, Kim2005}.
On the theoretical side, the similarities between the effective single-particle Hamiltonian for graphene in the low-energy regime and the Dirac equation for massless particles have been noted~\cite{Novoselov2005, Katsnelson2006, Geim2007} and lead to the prediction of ``relativistic effects'' in graphene.
One example is the non-transient {\it zitterbewegung} in the presence of an external magnetic field~\cite{rusin:125419, schliemann08}.
In order to study the {\it zitterbewegung} and other transport effects in graphene one needs to solve the time-dependent Dirac equation
with high accuracy and efficiency.
Previous time-dependent propagation methods have often been restricted to eigenstate decompositions.
However, the evaluation of integrals involving a finite set of eigenfunctions is cumbersome and also limited to energetically low lying eigenstates.
Additionally the eigenstate approach can only be used if the analytic solutions are known,
since the numerical determination of a full set of energy eigenstates is in general not feasible.

In section \ref{wppropagation} we show that the prior knowledge of the eigenstates is not required in order to obtain a reliable 
solution of the time-dependent Dirac-Hamiltonian of graphene.
Our algorithm is accurate up to the machine error and stable for any propagation time.
In addition it is very flexible and thus allows us to calculate the time-evolution in inhomogeneous magnetic fields and
arbitrary potential landscapes. In section \ref{wprevivals} we study quantitatively the scattering in realistic potentials and obtain
the fidelity (which is related to the mobility) of graphene in a magnetic field.
For wave-packets with large initial momenta we reveal a rich collapse and revival structure, 
which has a strong similarity to the dynamics of highly excited states of Rydberg atoms~\cite{PhysRevLett.56.716, PhysRevLett.61.2099}.
We derive analytic results for the semiclassical expectation value of the position operator
and the autocorrelation function which are in very good agreement with the numerical results.
Subsequently, we study the effect of an impurity potential and analyse the
suppression of the autocorrelation function corresponding to a rapid
fidelity decay.
Interestingly, revivals in the centre-of-mass coordinate survive the disorder-perturbations if the correlation
length of the potential is on the order of the cyclotron radii of the occupied states.
In general, our algorithm provides a basic method to model time-dependent
transport through mesoscopic systems~\cite{Kramer08, Aidala07, Stern08}.

\section{Wave-packet propagation for the graphene-system}
\label{wppropagation}

The electronic structure of graphene is obtained from the tight binding model of the two-dimensional honeycomb lattice.
In the resulting band structure two inequivalent vanishing bandgaps (often referred to as different valleys or Dirac points) appear, which are located at the corners of the Brillouin zone~\cite{neto:109, PhysRev.71.622}. 
The band structure around each Dirac point is conveniently described in the framework of an effective $\myvec k \cdot \myvec p$ perturbation theory and the
resulting Hamiltonian~\cite{PhysRevLett.53.2449}
%
%
%%%%%%%%%%
\begin{equation}
  \hat \mathcal H_\tau = - \ci \hbar v 
  \left (
    \tau \sigma_x \hat k_x + \sigma_y \hat k_y
  \right )
\label{DiracHamil}
\end{equation}
%%%%%%%%%%
%
%
has the same structure as the $2+1$ dimensional Dirac Hamiltonian, 
where the velocity is given by $v \approx 10^6\,\mathrm{m}/\mathrm{s}$,
and $\sigma_x$, $\sigma_y$ denote the corresponding Pauli matrices~\cite{PhysRevB.29.1685}.
The linear approximation is valid for energies close to the Dirac points.
For higher energies the trigonal warping gets important~\cite{saito00}.
We neglect this effect here, since we only consider wave packets which can be decomposed into
eigenstates from $-0.7\,\mathrm{eV}$ to $0.7\,\mathrm{eV}$.
The Hamiltonians \eref{DiracHamil} at the two Dirac points are obtained by switching the sign of $\tau = \pm 1$.
Here, we study the bulk region of a sheet of graphene and therefore we can neglect effects of the boundaries
which may couple the valleys and induce symmetry breaking~\cite{wimmer:177207, wurm:056806}.

In the following we present a new algorithm for the propagation of a wave-packet on a sheet of graphene
in an anisotropic magnetic field $B_z(\myvec r)$ oriented perpendicularly to the sheet. The Hamiltonian can 
also contain a position-dependent potential landscape $V(\myvec r)$.
The resulting system is described by the following Hamiltonian
%
%
%
%%%%%%%%%%
\begin{equation}
  \hat \mathcal H =
  \left (
        \begin{array}{c c}
%	V(\myvec r) & 	-\ci \hbar \hat k_x + \hbar \hat k_y - e \tilde A(\myvec r)
	V(\myvec r) & 	-\ci v \hbar \hat k_x + v \hbar \hat  k_y - v e  \tilde A(\myvec r)
	\\
%	-\ci \hbar \hat k_x + \hbar \hat k_y + e \tilde A^*(\myvec r) & V(\myvec r)
	-\ci v \hbar  \hat k_x - v \hbar \hat k_y - v e \tilde A^*(\myvec r) & V(\myvec r)
     \end{array}
  \right )
  \mathrm{.}
\end{equation}
%%%%%%%%%%
%
%
For simplicity we only use the Hamiltonian with $\tau = 1$, but it is straightforward
to apply the same algorithm to the second Dirac point.
The vector potential of the magnetic field $\myvec A(\myvec r) = (A_x(\myvec r), A_y(\myvec r),0)$ is 
connected to the magnetic field  $B(\myvec r) = \nabla \times \myvec A(\myvec r)$
and enters the Hamiltonian as $\tilde A(\myvec r) = A_x(\myvec r)- \ci A_y(\myvec r)$.
As initial state of the system we choose an arbitrary two spinor wavefunction
$\psi(t=0)  = \left\{ \psi_\mathrm{A}(t=0), \psi_\mathrm{B}(t=0)\right\}$.
The action of the time-evolution operator on the wavefunction is given by
%
%
%%%%%%%%%%
\begin{equation}
 | \psi (t) \rangle = \exp \left (- \frac{\ci}{\hbar} \hat \mathcal H t \right ) | \psi (0) \rangle
  \mathrm{.}
\end{equation}
%%%%%%%%%%
%
%
In principle, there exist several possibilities to calculate the action of the time-evolution operator. 
For a system with a discrete set of eigenstates defined by 
$\hat \mathcal H | \phi_n \rangle = \mathcal E_n | \phi_n \rangle$
a favourable way is given by the expansion in eigenstates.
In this case, the time-evolution becomes
%
%
%%%%%%%%%%
\begin{equation}
 | \psi (t) \rangle = \sum_n \langle \phi_n | \psi (0) \rangle
       \exp \left (- \frac{\ci}{\hbar} \mathcal E_n t \right ) | \phi_n \rangle
  \mathrm{.}
  \label{eigenstate_time_evolution}
\end{equation}
%%%%%%%%%%
%
%
Such a propagation algorithm was already used in references~\cite{rusin:125419, schliemann08} to evaluate
the {\it zitterbewegung} in graphene and also for other semiconductor materials~\cite{schliemann:125303}.
Due to the lack of analytical known solutions of the Dirac equation, this scheme can only be applied to a limited set of problems.
Thus it is favourable to develop an algorithm which does not require any knowledge of the eigenstates
and can be applied to arbitrary mesoscopic systems.
A similar starting point exists in quantum chemistry, where no analytic solutions are known for
the molecular potential energy surfaces.
In all these cases it is favourable to use an algorithm which directly gives the time-evolution
of the system without evaluating all eigenstates first~\cite{Feit1982412}.
All relevant information about stationary properties, i.e.\ the energy spectrum or the Green's function and
the local density of states, is encoded in time-dependent observables, which are extracted afterwards
from the time-evolved wave-packet~\cite{Heller81, Kramer08}.

For the time-dependent solution of the Dirac Hamiltonian, we use a reliable and efficient algorithm
based upon the polynomial expansion of the propagator in Chebyshev polynomials~\cite{tal-ezer:3967}.
The main idea of the algorithm is to expand the time-evolution operator on the interval
$[-1,1]$ using a set of Chebyshev polynomials, which are particularly suitable since they
distribute the numerical error of the expansion equally on the whole interval.
As a first step the eigenvalues of the Hamiltonian $\hat \mathcal H$ have to be projected
into the interval of the expansion by introducing a normalised Hamiltonian
%
%
%%%%%%%%%%
\begin{equation}
 \hat H_\mathrm{norm} = 2 \frac{ \hat \mathcal H}{\Delta E}\mathrm{.}
  \label{scaledHamil}
\end{equation}
%%%%%%%%%%
%
%
The scaling-energy $\Delta E$ has to be big enough to cover all eigenenergies
contributing to the initial wave-packet.
With the aid of \eref{scaledHamil} the time-evolution is given by
%
%
%%%%%%%%%%
\begin{equation}
| \psi (t) \rangle  =  \sum_{n=0}^N a_n \left ( \frac{\Delta E t}{2 \hbar} \right)
\underbrace{T_n (- \ci \hat H_\mathrm{norm}) | \psi (0) \rangle }_{P_n}
\mathrm{,}
\end{equation}
%%%%%%%%%%
%
%
where $T_n$ stands for the series of Chebyshev polynomials of the first kind.
The expansion coefficients are given by
%
%
%%%%%%%%%%
\begin{equation}
a_n  \left ( \frac{\Delta E t}{2 \hbar} \right) =  (2-\delta_{n,0}) \ J_n\left (
\frac{\Delta E t}{2 \hbar} \right)
\mathrm{,}
\end{equation}
%%%%%%%%%%
%
%
with the Bessel function $J_n$.
Only the action of the polynomial of $\hat H_\mathrm{norm}$ on the initial state is needed
to calculate the time-evolution.
Thus it is sufficient to recursively calculate the vectors $P_n$ by the
following modified recursion relations for Chebyshev polynomials:
%
%
%%%%%%%%%%
%\begin{subequations}
\begin{eqnarray}
  P_0 &=&  | \psi (0) \rangle  \mathrm{,}\\
  P_1 &=&  \hat H_\mathrm{norm} | \psi (0) \rangle  \mathrm{,}\\
  P_n &=&  - 2 \ci \hat H_\mathrm{norm}  P_{n-1} +  P_{n-2} \mathrm{.}
\end{eqnarray}
%\end{subequations}
%%%%%%%%%%
%
%
In order to avoid the fermion doubling problem of a simple first-order nearest-neighbour
derivative~\cite{Nielsen198120} we use a Fourier representation of the spinors and apply 
the kinetic energy operator in momentum space~\cite{Feit1982412}.
Another possibility is to introduce certain non-local first-order derivatives~\cite{o:235438}.
In our case, the action of the normalised Hamiltonian acting on a two spinor wavefunction
$\left( \psi_\mathrm{A}, \psi_\mathrm{B}\right)$
is given by
%
%
%%%%%%%%%%
\begin{equation*}
 \hat H_\mathrm{norm} \left (
    \begin{array}{c}
	\psi_\mathrm{A}
	\\
	\psi_\mathrm{B}
     \end{array} 	\right )
 = \frac{2}{\Delta E} \left (
    \begin{array}{c}
	V(\myvec r) \psi_\mathrm{A} 
	- v \left (\hbar \mathcal F^{-1} \left ( \ci k_x - k_y \right ) \mathcal F
	+ e \tilde A(\myvec r) \right ) \psi_\mathrm{B}   
	\\
	V(\myvec r) \psi_\mathrm{B}
	- v \left (\hbar \mathcal F^{-1} \left ( \ci k_x + k_y \right ) \mathcal F
	+ e \tilde A^*(\myvec r) \right ) \psi_\mathrm{A}   
     \end{array} 	\right )
\end{equation*}
%%%%%%%%%%
%
%
where $\mathcal F^{-1}$ and $\mathcal F$ stand for the two-dimensional Fast Fourier Transform.
This type of propagation algorithm allows us to evaluate the time-evolution of the wave-packet for
arbitrary long propagation times, as long as the wavefunction does not leave the region of interest.
A good measure of the propagation quality is given by the comparison of the analytic autocorrelation function for the homogeneous magnetic field 
with the numerical result.
Our results show that the maximum error is indeed on the order of the unavoidable finite machine precision.
In contrast to all other propagation algorithms for graphene we are able to incorporate arbitrarily shaped
potentials and also inhomogeneous magnetic fields in our configurations.

\section{Wave-packet revivals}
\label{wprevivals}
\subsection{Model and observables}
In the following we apply the propagation algorithm to an infinite sheet of graphene put in
a perpendicular magnetic field $\myvec B(\myvec r) = (0,0,B)$ and set the potential $V(\myvec r)$
to zero.
This system was proposed by Rusin {\it et. al.}~\cite{rusin:125419} and Schliemann~\cite{schliemann08}
to study the non-transient {\it zitterbewegung} of a wave-packet on graphene.
The eigenstates of the Dirac Hamiltonian in a perpendicular magnetic field are known and given by
the harmonic oscillator eigenfunctions~\cite{arxiv:0811.4595}.
In the lowest Landau level, the ground state wavefunction has Gaussian shape and is located on one sublattice.
As initial condition we consider a kicked Gaussian of the form
%
%
%%%%%%%%%%
\begin{equation}
 | \psi (\myvec r, t=0) \rangle = 
   \frac{1}{\sqrt\pi a_0} \exp \left ( -\frac{ (\myvec r - \myvec r_0)^2}{2 a_0^2} + \ci \myvec k \myvec r \right )
   \left (
    \begin{array}{c}
	1
	\\
	\e^{\ci \varphi}
     \end{array}
    \right )
   \mathrm{.}
	\label{initial_wp}
\end{equation}
%%%%%%%%%%
%
%
The width is taken as $a_0 = \hbar/(eB)$ and the initial momentum
$\myvec k$ can point along any direction in the $x-y$ plane, 
but is set to $\myvec k = k_0 \myvec e_y$ in the following.
The pseudospin $\vec \sigma \myvec k$ of the wave-packet is determined by the phase-relation
between the two spinor entries.
For $\varphi = \pi/2$, the wave-packet is mostly electron-like since the momentum is pointing along the
$y$-direction.
Consequently the wave-packet is mostly hole-like if $\varphi = 3\pi/2$ and a
mixture of both states for intermediate angles.

\begin{figure}[t]
  \centering
  \begin{tabular}[t]{p{.3cm}cc}
       & $\myvec k \parallel \vec \sigma \myvec k$, $\varphi = \pi/2$
       & $\myvec k \perp \vec \sigma \myvec k$, $\varphi = \pi$\\	
  \vspace*{-1.7cm}(a)\vspace*{1.0cm}
       & \hspace*{0.018\columnwidth}
         \includegraphics[width=0.325\columnwidth]{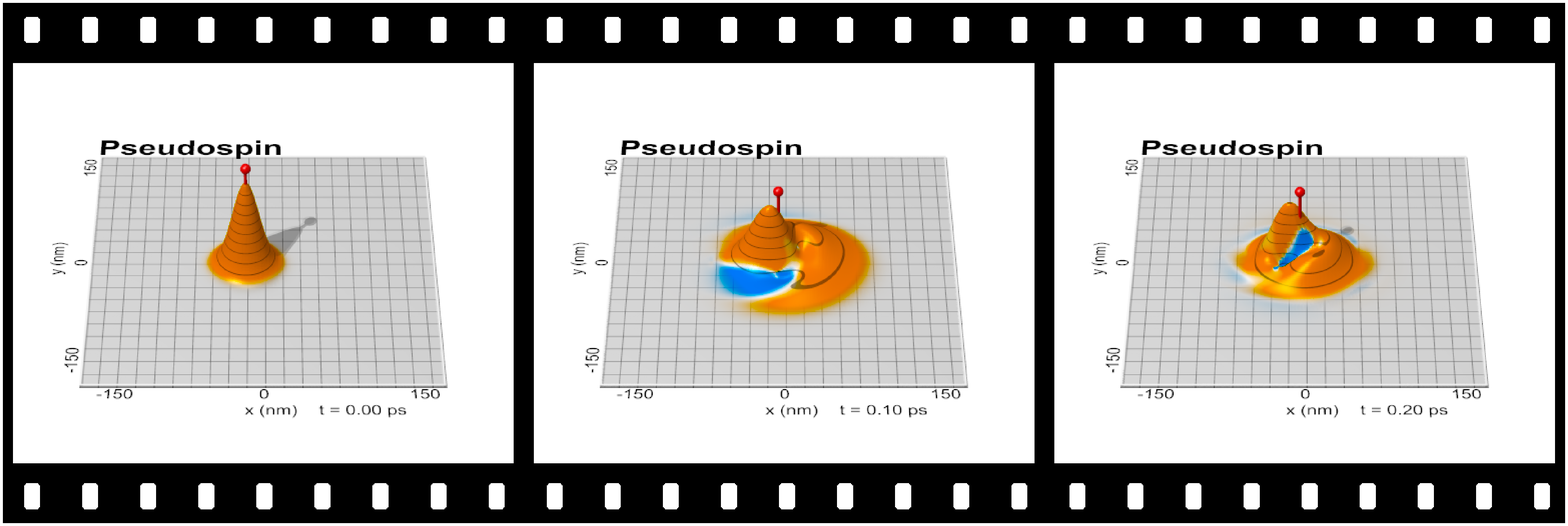}
       & \hspace*{0.014\columnwidth}
         \includegraphics[width=0.325\columnwidth]{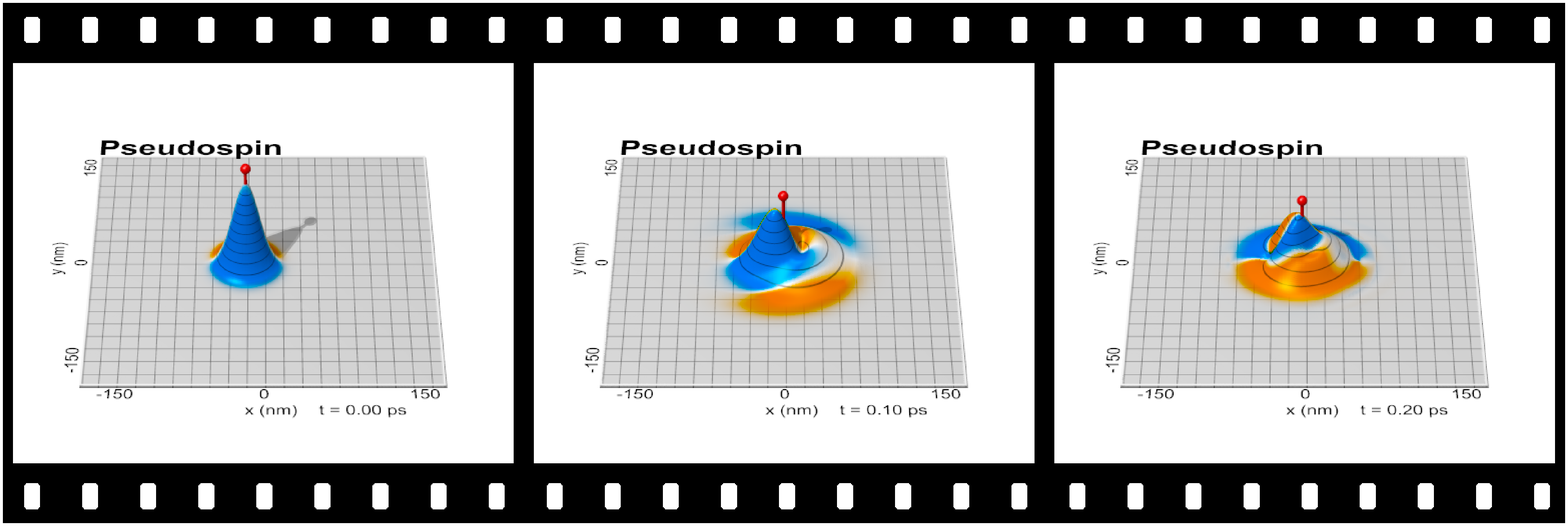} \\
   \vspace*{-6.0cm}(b)\vspace*{5.3cm}
       & \multicolumn{2}{l}{	
         \includegraphics[width=0.75\columnwidth]{figure1b.eps}
        }
  \end{tabular}
        \caption{
	(a) Some frames of the time-evolution of a kicked Gaussian wave-packet for different initial
	pseudospin configurations ($B = 5\,\mathrm{T}$,	$k_0 = \sqrt{eB/\hbar}$, $a_0=\hbar/(eB)$).
	The red pin stands for the centre-of-mass, the height for the absolute value and the
	colour for the local pseudospin given by
	$\langle \psi(\myvec r, t) | \vec \sigma \hat \myvec k | \psi(\myvec r, t) \rangle$.
	A video of the propagation is available from \url{http://www.quantumdynamics.de/graphene} 
	for initial pseudospin $\vec \sigma \myvec k \parallel \myvec k$ ({\tt movie1a.mp4}, $3.0\,\mathrm{MB}$)
	and $\vec \sigma \myvec k \perp \myvec k$ ({\tt movie1b.mp4}, $3.2\,\mathrm{MB}$).
	(b) Local density of states calculated from the wave-packet  propagation.
	The spectrum is antisymmetric if the initial pseudospin is parallel to the momentum and 
	symmetric if pseudospin and momentum are orthogonal to each other.
	Red dots denote the analytical energies of the Landau levels and coincide up to a 
	very high accuracy with the numerical positions.
	\label{lowkick_LDOS}
	}
\end{figure}

During the propagation we track several observables, including the expectation value of the position operator
%
%
%%%%%%%%%%
\begin{equation}
  \myvec r(t) =
   \langle \psi(t) | \hat{\myvec r} | \psi(t) \rangle 
\mathrm{.}
\end{equation}
%%%%%%%%%%
%
%
We refer to $\myvec r(t)$ as the ``centre-of-mass'' although the calculated
motion describes massless particles.
The trembling motion of the centre-of-mass is visible in the
videos linked from \fref{lowkick_LDOS}(a) for different initial setups.
Another important property, the local pseudospin, is given by 
$\langle \psi(\myvec r, t) | \vec \sigma \hat \myvec k | \psi(\myvec r, t) \rangle$
at each gridpoint $\myvec r$ and colour coded in the video.
The local pseudospin reveals the local electron or hole character of the
wave-packet and shows an anti-symmetry with respect to the $x$-axis if the 
pseudospin of the initial state is oriented orthogonal to the initial momentum.
%
%
%%%%%%%%%
%%%%%%%%%%%%%%%
%
%
An advantage of our method is the extraction of stationary properties, like
the local density of states (LDOS) and the Green's function, out of
the time-evolution~\cite{Kramer08}.
For example the relative strength of each energy eigenstate with respect 
to the initial state is given by the Fourier transform of the autocorrelation function
%
%
%%%%%%%%%%
\begin{equation}
  C(t) = \langle \psi(t=0) | \psi(t) \rangle
  \mathrm{.}
\end{equation}
%%%%%%%%%%
%
%
The same concept allows us to calculate the $\myvec E \times \myvec B$ drift motion and eigenstates in the
quantum Hall regime of graphene for high bias currents~\cite{arxiv:0811.4595}.
The LDOS is given by the Laplace transformation of the
autocorrelation function
%
%
%%%%%%%%%%
\begin{equation}
  n(E) \propto - \Im \int_0^\infty dt\ \e^{\ci E t/\hbar} C(t)
  \mathrm{.}
\end{equation}
%%%%%%%%%%
%
%
As shown in \fref{lowkick_LDOS}(b) the LDOS of the trembling wave-packet
reveals pronounced peaks at the energies of the Landau levels
$\mathcal E_n = v~\mathrm{sign}(n) \sqrt{2 e B \hbar |n|}$.
The peak height (which is proportional to the enclosed area in the LDOS) indicates the overlap of 
the initial Gaussian wave-packet with the different
Hermite polynomials of the Landau levels.
For $\varphi=\pi/2$, mostly electrons are occupied whereas for
$\varphi=\pi$ holes and electrons are equally occupied.
Our computational scheme derives the overlap information without calculating the 
numerically unstable integrals over Hermite polynomials used in previous methods.
Therefore we can increase the initial momentum $k_0$ of the wave-packet and study the
quantum mechanical propagation of high Landau levels.

\begin{figure}[t]
 \centering \includegraphics[width=0.75\columnwidth]{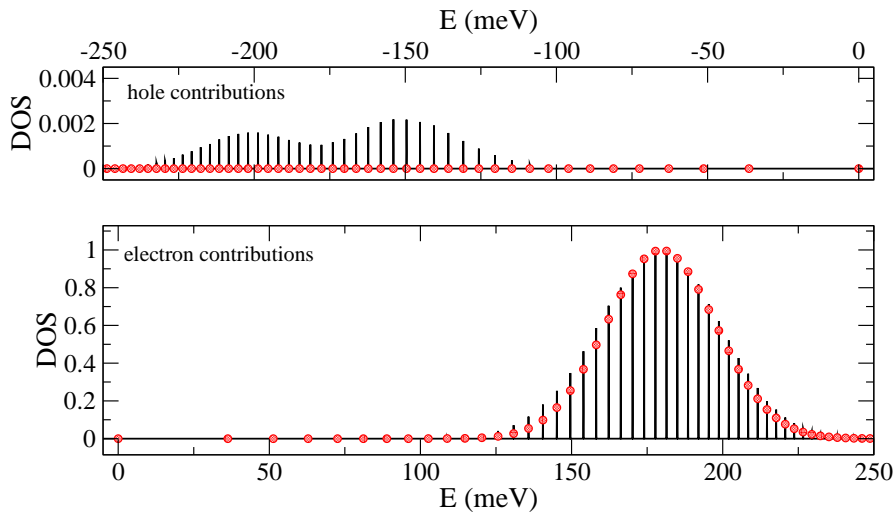}
        \caption{
	Spectrum of a kicked Gaussian wave-packet with an average occupied Landau level of $\bar n = 24.5$
	($B = 1T$, $k_0 = 7\sqrt{eB/\hbar}$).
	The hole-like contributions are approximately 400 times smaller than the electron-like parts.
	The red dots show the position of the analytical Landau levels as well as the probability distribution
	\eref{gaussdist} used for the following approximations.
	\label{highkick_LDOS}
	}
\end{figure}

\begin{figure}[t]
 \centering \includegraphics[width=0.75\columnwidth]{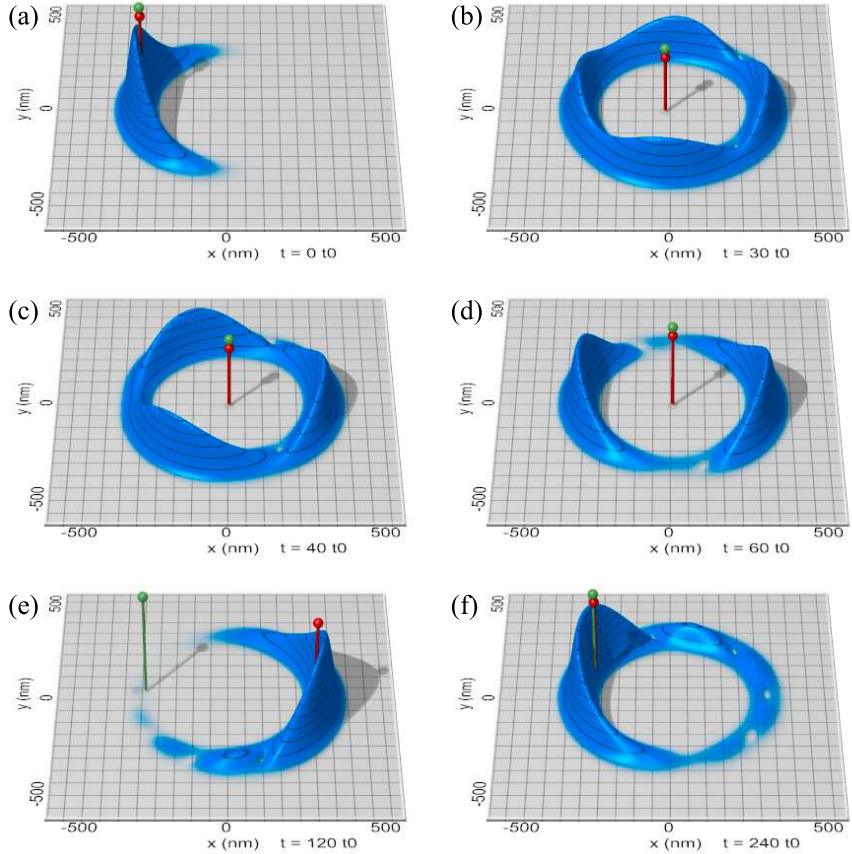}
        \caption{Revivals and fractional revivals of a cyclotron wave-packet in a magnetic
	field of $B=1\,\mathrm{T}$.
	The red pin stands for the quantum mechanical centre-of-mass, whereas the
        green pin is calculated with the semiclassical expression \eref{CMclassical}.	
	The initial wave-packet in (a) was chosen such that the contributing eigenstates
	follow the Gaussian distribution \eref{gaussdist} with $\bar n = 60$, $\sigma = 1$, leading to a
	Poincar\'e cyclotron time of $t0=T_\mathrm{cl}$.
	During the time-evolution the centre-of-mass collapses to the centre of a circular
	wavefunction and later shows pronounced revivals. For example
	(b) displays a quarter revival at $t= 1/2 \bar n\,T_\mathrm{cl}$,
	(c) a third revival at $t= 2/3 \bar n\,T_\mathrm{cl}$ and
	(d) a half revival at $t= \bar n\,T_\mathrm{cl}$.
	All revivals have specific symmetries which also occur in the
	subsequent revivals.
	After a certain time a localised wave-packet emerges in 
	(e) as a mirror revival at $t= 2 \bar n\,T_\mathrm{cl}$.
	In this case, the quantum mechanical centre-of-mass is at the opposite side
        compared to the semiclassical centre-of-mass. This phenomenon has already been seen in
	Rydberg revivals~\cite{PhysRevA.42.6308}.
	(f) The full revival where the semiclassical and the quantum mechanical centre-of-mass
        coincide occurs for $t= 4 \bar n\,T_\mathrm{cl}$.
	The related movie ({\tt movie2.mp4}, $3.0\,\mathrm{MB}$) which illustrates the dynamics is available from
        \url{http://www.quantumdynamics.de/graphene}.
	\label{revivals}
	}
\end{figure}

The trembling motion on graphene is present since both, electron and hole-like states,
contribute to the initial wave-packet.
For high initial momenta the index of the average Landau level is given by
%
%
%%%%%%%%%%
\begin{equation}
 \bar n = \frac{\hbar}{2 e B} k_0^2 \mathrm{,}
\end{equation}
%%%%%%%%%%
%
%
which follows from the semiclassical quantisation condition \eref{quantization_condition}.
Our quantum mechanical calculations support this assertion as shown in the LDOS of \fref{highkick_LDOS}.
%
%
%%%%%%%%%
%%%%%%%%%%%%%%%
%
%
A strongly kicked Gaussian wave-packet mostly occupies one energy branch
if the initial pseudospin is parallel to the initial momentum.
For an initial momentum which puts the Gaussian wave-packet into the
$24.5^{\mathrm{th}}$ Landau level, the hole contributions are 400 times smaller compared to the electron parts.
In such a scenario the trembling motion may be neglected and other interesting revival phenomena appear.
In atomic physics revival effects are commonly observed 
for valence electrons in the Coulomb potential of Rydberg atoms
and have been studied in theory~\cite{PhysRevLett.56.716} as well as in
experiments~\cite{PhysRevLett.61.2099, PhysRevLett.64.2007}.
Reference~\cite{Robinett20041} contains a recent review about revivals of quantum wave-packets.
In order to get an idea about the rich revival structure of a quantum wave-packet on graphene,
it is instructing to look at the numerical results in \fref{revivals} and the corresponding video.
Wave-packet revivals are typically not present in semiconducting material in a perpendicular
magnetic field since there the dynamics is governed by a purely quadratic Hamiltonian.
Graphene and toplogical insulators form their own class due to their
non-quadratic Hamilton operators.
%
% Revivals and fractional revivals of a cyclotron wave-packet
%%%%%%%%%
%%%%%%%%%%%%%%%
%
%

\subsection{Revivals in the centre-of-mass position}

As first phenomena we study the collapse and the revival of the centre-of-mass of a graphene wave-packet.
The quantum mechanical position of the centre-of-mass cannot be found analytically.
Thus we introduce a classical picture where the centre-of-mass is given by a weighted sum over the
centre-of-masses moving along quantised cyclotron orbits.
The centre-of-mass of the classical cyclotron motion in the $n$-th Landau level is defined by
$z(t) = x(t)+ \mathrm{i} y(t)$
with
%
%
%%%%%%%%%%
\begin{equation}
 z_n(t) = l_n e^{\mathrm{i} \omega_n  t}
\end{equation}
%%%%%%%%%%
%
%
as derived in \ref{appendixA}.
The occupation $P_n$ depends on the initial momentum $k_0$, the initial pseudospin
and the width $a_0$ of the wave-packet.
As an approximation we find from numerical calculations for high average Landau levels a Gaussian
distribution of the form
%
%
%%%%%%%%%%
\begin{equation}
 P(n) = \frac{1}{\sqrt{2 \pi} \sigma} \exp \left( -\frac{(n-\bar n)^2}{2 \sigma^2} \right),
 \label{gaussdist}
\end{equation}
%%%%%%%%%%
%
%
shown in \fref{highkick_LDOS}.
Here $\bar n$ stands for the average occupied Landau level and $\sigma$ denotes the Landau level spread.
The sum over the Landau levels is normalised if $\bar n \gg \sigma$ is fulfilled.
In this limit the cyclotron radii $l_n$ do not differ substantially from the cyclotron radius of the average occupied
Landau level $l := l_{\bar n}$.
Thus, the  time-evolution of the average position is approximated by
%
%
%%%%%%%%%%
\begin{equation}
 z(t) = l \sum_{n=-\infty}^{\infty} P_n e^{\mathrm{i} \omega_n  t}
\mathrm{.}
\label{CMclassical}
\end{equation}
%%%%%%%%%%
%
%
The last sum contains both, the centre-of-mass collapse as well as the revivals.
Numerical results of this classical centre-of-mass motion are shown in \fref{revivals}.
In the corresponding video the classical approximation is marked by the green pin, which 
almost perfectly coincides with the exact quantum mechanical result indicated by the red pin.

\subsubsection[Collapse]{Collapse:}
%
% Collapse of a Dirac wave-packet
%%%%%%%%%
\begin{figure}[t]
 \centering \includegraphics[width=0.75\columnwidth]{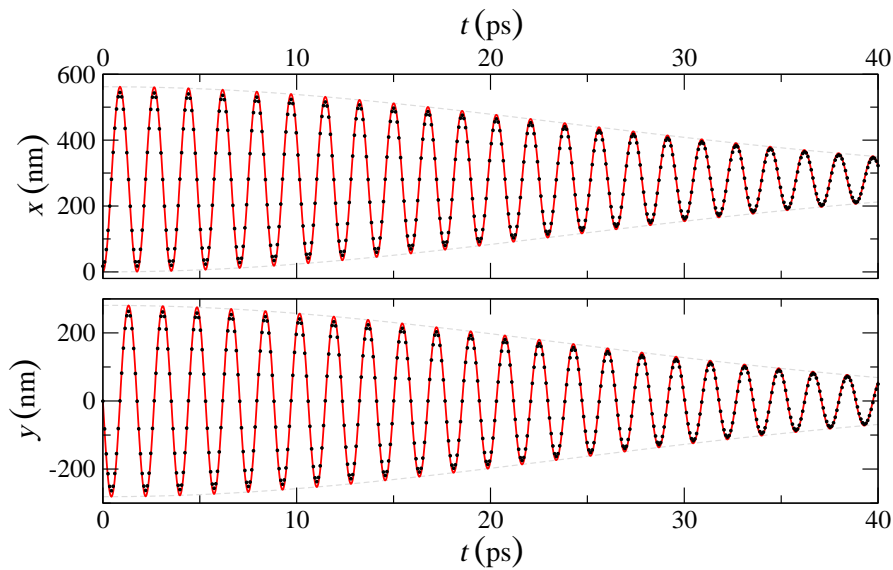}
   \caption{Collapse of a Dirac wave-packet ($\bar n = 60$, $\sigma = 1$, $B=1\,\mathrm{T}$).
	Both panels show the time-evolution of the $x$- and the $y$-position of the wave packet.
	The black dots are obtained from a quantum mechanical calculation and fit
	quite well to the classical centre-of-mass which is shown by the red line
	and given by the semiclassical approximation \eref{simplecollapse}.
	A small deviation is visible which stems from the fact that the centre-of-mass of the initial
	wave-packet has already a strong angular dispersion (see \fref{revivals}) and
	thus the average cyclotron radius $l$ is overestimated.
	\label{fig_collapse}
	}
\end{figure}
%%%%%%%%%%%%%%%
%
%
Next, we extract the times of the initial collapse of the centre-of-mass and the revivals out of equation
\eref{CMclassical}. We expand the cyclotron frequency  $\omega_n$ around the average Landau level $\bar n$:
%
%%%%%%%%%%
\begin{equation}
 \omega(n) \approx  v \sqrt{\frac{e B}{2 \hbar}} \bar n^{-1/2} \left (
  1 - \frac{1}{2} \frac{n-\bar n}{\bar n} + \frac{3}{8} \frac{(n - \bar n)^2}{\bar n^2} - ...
 \right )
  \mathrm{.}
\end{equation}
%%%%%%%%%%
%
%
For high Landau levels the energy spectrum becomes denser and the sum can be approximated by an
integral which we evaluate analytically
after the series is reduced to the first two terms:
%
%
%%%%%%%%%%
\begin{eqnarray}
 z(t) &=& l \int_{-\infty}^{\infty} dn P(n) e^{\mathrm{i} \omega(n)  t} \nonumber\\
      &\approx& \frac{l}{\sqrt{2 \pi} \sigma}
          \int_{-\infty}^{\infty} dn
          \exp \left( -\frac{(n-\bar n)^2}{2 \sigma^2} \right)
          \exp \left( \mathrm{i} v \sqrt{\frac{e B}{2 \hbar}} \frac{1}{\sqrt{\bar n}}
             \left ( 1 - \frac{1}{2} \frac{n-\bar n}{\bar n} \right ) t
          \right) \nonumber\\
%       &=& \frac{k_0 l^2}{\sqrt{2 \pi} \sigma}
%            \exp \left( \mathrm{i} v \sqrt{\frac{e B}{2 \hbar}} \bar n^{-1/2} t \right )
%            \exp \left( -\frac{e B v^2}{8 \hbar} \frac{\sigma^2 t^2}{\bar n^3} \right )
      &=& l \exp \left( \mathrm{i} \omega_{\bar n} t \right ) \cdot
            \exp \left( -\frac{e B v^2}{8 \hbar} \frac{\sigma^2 t^2}{\bar n^3} \right )
        \mathrm{.}
	\label{simplecollapse}
\end{eqnarray}
%%%%%%%%%%
%
%
Equation \eref{simplecollapse} shows that the centre-of-mass motion completes one cyclotron orbit
during the cyclotron time of the average Landau level $\bar n$
%
%
%%%%%%%%%%
\begin{equation}
  T_\mathrm{cl}= \frac{2 \pi}{v} \sqrt{\frac{2 \hbar}{e B} |\bar n|}
   \mathrm{,}
\end{equation}
%%%%%%%%%%
%
%
and also a Gaussian decay on the timescale
%
%
%%%%%%%%%%
\begin{equation}
%  T_\mathrm{coll}= 4 \sqrt{\frac{\hbar}{e B v^2}} \frac{\bar n^{3/2}}{\sigma}
  T_\mathrm{coll}= \frac{4}{v \sigma} \sqrt{\frac{\hbar}{e B} |\bar n|^3}
  \mathrm{.}
\end{equation}
%%%%%%%%%%
%
%
Our numerical calculations in \fref{fig_collapse} show the same behaviour.
Very small deviations from the exact result stem from the initial radial dispersion of the
Gaussian wave-packet, which is not included in the classical theory. 

\subsubsection[Revivals]{Revivals:}

The revival of the classical centre-of-mass takes place if all particles orbiting along the contributing 
Landau orbits reunite at the initial position.
We calculate the revival time from the Taylor expansion of the classical centre-of-mass motion inserted
in equation \eref{CMclassical}.
In order to meet the conditions for a revival $|z(  T_\mathrm{rev(CM)} )| = l$, all elements of the sum
%
%
%%%%%%%%%%
\begin{equation}
  z(t) = l \sum_{n=-\infty}^\infty P_n \exp \left (
    \ci \omega_{\bar n} t +     \ci (n-\bar n)\omega'_{\bar n} t + ...
 \right )
\end{equation}
%%%%%%%%%%
%
%
must have the same phase.
Thus, the $n$-dependent addend of the exponent $\ci n  \omega'_{\bar n} t$
must be a multiple of $2 \pi \ci$.
This leads to a revival time of the classical centre-of-mass given by
%
%
%%%%%%%%%%
\begin{equation}
  T_\mathrm{rev(CM)} 
     = \frac{2 \pi}{\omega'_{\bar n}} 
     = \frac{4 \pi}{v} \sqrt{\frac{2 \hbar}{e B} |\bar n|^3}
     = 2 |\bar n|\,T_\mathrm{cl}
  \mathrm{.}
\end{equation}
%%%%%%%%%%
%
%
%
%
% CM revival
%%%%%%%%%
\begin{figure}[t]
 \centering \includegraphics[width=0.75\columnwidth]{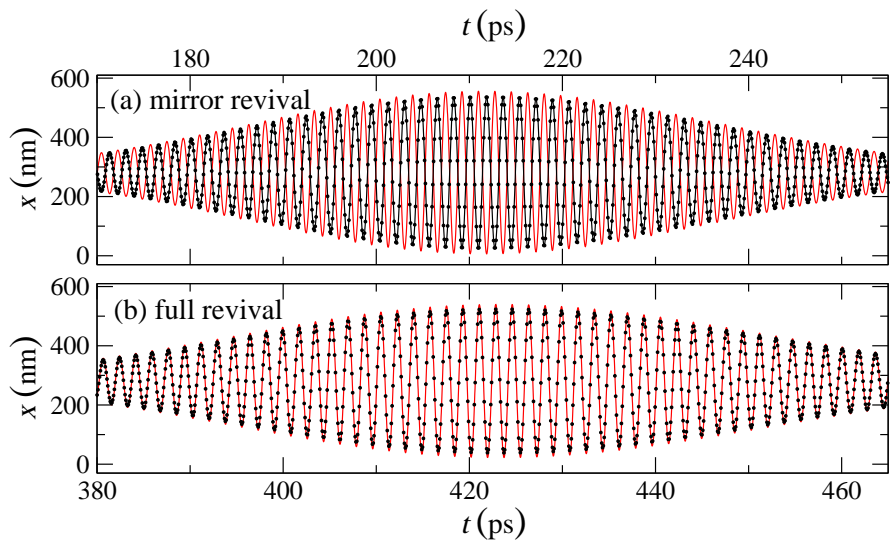}
   \caption{(a) First and (b) second revival of the centre-of-mass  ($\bar n = 60$, $\sigma = 1$, $B=1\,\mathrm{T}$).
	Red line shows the centre-of-mass given by the semiclassical approximation \eref{CMclassical}.
	This result is compared with a full quantum mechanical calculation
	shown by the black dots.
	Both results fit well with the remarkable difference for odd revivals. The quantum
	mechanical result is opposing the classical result, which has already been
	seen in Rydberg atoms.
	\label{fig_CM_revival}
	}
\end{figure}
%%%%%%%%%%%%%%%
%
%
In this case the classical centre-of-mass of the wave-packet is on the opposite side compared to the
centre-of-mass of the quantum mechanical calculation.
This phenomenon is visible in the comparison between the classically and the quantum mechanically
calculated expectation value of the $x$ coordinate for the first revival time
shown in \fref{fig_CM_revival}(a).
Exactly the same effect occurs in the mirror revivals of Rydberg atoms~\cite{PhysRevA.42.6308, PhysRevA.48.2574}.
For the second revival the classical and the quantum mechanical centre-of-mass coincide again
as is evident from \fref{fig_CM_revival}(b).

\subsection{Revivals in the autocorrelation function}

In contrast to the centre-of-mass revivals, the revivals in the autocorrelation function
can be understood without a semiclassical approximation.
By definition the time-evolution of the autocorrelation function is given by
%
%
%%%%%%%%%%
\begin{equation}
  C(t) = \langle \phi(t=0) | \phi(t) \rangle = \sum_{n=-\infty}^\infty \e^{-\ci \mathcal E_n t/\hbar} P(n)
	\mathrm{.}
\end{equation}
%%%%%%%%%%
%
%
Again, we expand the square root dependence in a Taylor series
%
%
%%%%%%%%%%
\begin{equation}
  \mathcal E(n) = v  \sqrt{2 e B \hbar |\bar n|}
   \left (
	1+ \frac{1}{2}\frac{n-\bar n}{\bar n}-\frac{1}{8}\frac{(n-\bar n)^2}{\bar n^2}
	+\frac{1}{16}\frac{(n-\bar n)^3}{\bar n^3} +...
   \right )
  \mathrm{,}
\end{equation}
%%%%%%%%%%
%
%
leading to an approximate autocorrelation function
%
%
%%%%%%%%%%
\begin{equation}\label{eq:ACapp}
  C(t) =  \sum_{n=-\infty}^\infty \e^{-\ci \left (
      \mathcal E_{\bar n} 
    + (n-\bar n) \mathcal E'_{\bar n}
    + \frac{1}{2}(n-\bar n)^2 \mathcal E''_{\bar n} 
    + \frac{1}{6}(n-\bar n)^3 \mathcal E'''_{\bar n}
    + ...
   \right ) t/\hbar} P(n)
   \label{approxC}
   \mathrm{,}
\end{equation}
%%%%%%%%%%
%
% 
where $\mathcal E'_{\bar n}$ stands for the first derivative of the dispersion 
and $\mathcal E''_{\bar n}$, $\mathcal E'''_{\bar n}$  denote the higher order derivatives.
It is astonishing that all orders are prominently encoded in the quantum mechanical motion
and influence the picture at the relevant revival times.
%
% different oscillation frequencies
%%%%%%%%%
\begin{figure}[t]
 \centering \includegraphics[width=0.75\columnwidth]{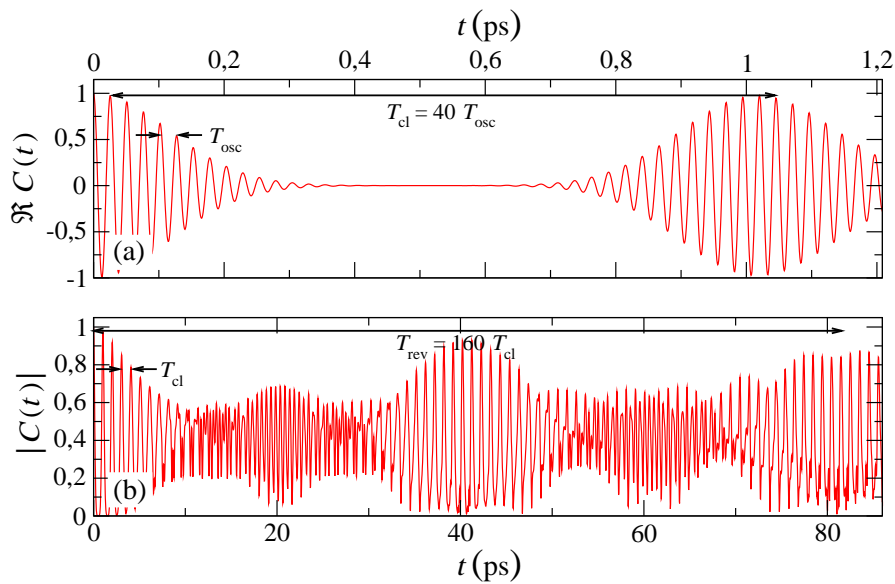}
   \caption{Revivals in the autocorrelation function of a graphene cyclotron wave-packet 
	($\bar n = 20$ $\sigma = 1$, $B = 1\,\mathrm{T}$).
	(a) The real part shows fast oscillations on the timescale $T_\mathrm{osc}$
	which exhibit revivals every $T_\mathrm{cl}$ corresponding to the time of
	a classical cyclotron orbit.
	During the time in panel (b) the dispersion of the initially localised wave-packet
	grows, leading to a uniform distribution which recovers to a full revival
	after $T_\mathrm{rev}$.
	\label{fig_fluc}
	}
\end{figure}
%%%%%%%%%%%%%%%
%
%
From the term $\ci \mathcal E_{\bar n}  t/\hbar$ in the exponent of equation \eref{eq:ACapp} we obtain the phase oscillation of the autocorrelation function
on the timescale
%
%
%%%%%%%%%%
\begin{equation}
  T_\mathrm{osc} 
     = \frac{2 \pi \hbar}{|\mathcal E_{\bar n}|} = \frac{\pi}{v} \sqrt\frac{2 \hbar}{e B |\bar n|}
\mathrm{.}
\end{equation}
%%%%%%%%%%
%
%
This oscillation takes place when the wave-packet moves over the initial position.
At the classical cyclotron time
%
%
%%%%%%%%%%
\begin{equation}
  T_\mathrm{cl} 
     = \frac{2 \pi \hbar}{|\mathcal E'_{\bar n}|}
     = \frac{2 \pi}{v} \sqrt\frac{2 \hbar | \bar n|}{e B}
    \mathrm{,}
\end{equation}
%%%%%%%%%%
%
%
the wave-packet comes back to the initial position and thus the autocorrelation function
approximately reaches its initial value, as shown in \fref{fig_fluc}(a).
For a revival the terms proportional to the second derivative have to be a multiple of 
$2 \pi$.
This leads to a revival for
%
%
%%%%%%%%%%
\begin{equation}
  T_\mathrm{rev} 
     = 2 \frac{2 \pi \hbar}{|\mathcal E''_{\bar n}|}
     = \frac{8 \pi}{v} \sqrt\frac{2 \hbar |\bar n|^3}{e B}
     = 4 |\bar n|\,T_\mathrm{cl} 
     \mathrm{,}
\end{equation}
%%%%%%%%%%
%
%
which is best seen in the time-evolution of the absolute value of the autocorrelation function
(\fref{fig_fluc}(b)).
Of course, the revival hierarchy continues to higher orders.
In atomic physics the next time is the so called ``super revival time''
%
%
%%%%%%%%%%
\begin{equation}
  T_\mathrm{super} 
     = 6 \frac{2 \pi \hbar}{|\mathcal E'''_{\bar n}|}
     = \frac{16 \pi}{v} \sqrt\frac{2 \hbar |\bar n|^5}{e B}
     = 8 \bar n^2\,T_\mathrm{cl} 
     = 2 |\bar n|\,T_\mathrm{rev}
\mathrm{.}
\end{equation}
%%%%%%%%%%
%
%
The huge range of timescales requires a highly accurate wave-packet propagation, since the
phase of the propagated wave-packet needs to be accurate for at least $4 |\bar n|^2$
phase oscillations in the autocorrelation function.
The used polynomial propagation is well suited for this task since the accumulated error
is reduced by using only very few but long timesteps.

\subsection{Fractional revivals}
%
% Poincare autocorrelation
%%%%%%%%%
\begin{figure}[t]
 \centering \includegraphics[width=0.75\columnwidth]{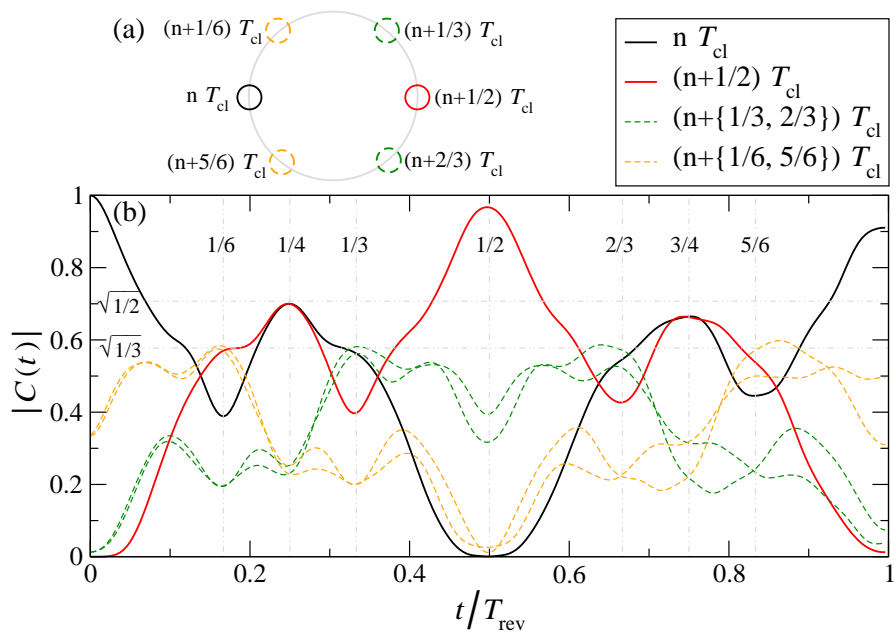}
   \caption{Poincar\'{e} sections of the autocorrelation function in order to study fractional
	revivals ($\bar n = 60$, $\sigma = 1$, $B=\,\mathrm{T}$).
	The autocorrelation function is calculated for successive classical cyclotron times $n$ and suitable subdivisions 
	$p/q$ as shown in (a).
	(b) The resulting autocorrelation function shows full revivals, a mirror revival and several fractional revivals.
	There the absolute value features local extremal points which are labelled in the plot. 
	\label{fig_fractional}
	}
\end{figure}
%%%%%%%%%%%%%%%
%
%
Interestingly, we can also distinguish fractional revivals occurring between the full revivals~\cite{Averbukh1989449}
which are again encoded in the autocorrelation function~\cite{PhysRevLett.77.3999}.
Fractional revivals have been experimentally measured in Rydberg atoms for the first time~\cite{PhysRevLett.72.3783}.
The underlying phase effect can be also employed for prime number factorisation~\cite{merkel2006, stefanek2007}.
Recently several experiments have implemented this scheme
into a NMR setup~\cite{mehring:120502}, cold atoms~\cite{gilowski:030201} or optically~\cite{bigourd:030202}
and successfully factorised numbers. 
Note that the phase properties for wave-packets in graphene differ from those of the aforementioned
experiments.

In order to extract the fractional revivals one has to calculate the autocorrelation function in a Poincar\'{e} map
for classical cyclotron time differences.
\Fref{fig_fractional} displays the fractional revivals for several $t = (n+p/q)\,T_\mathrm{cl}$
where $n$ is an integer and $p/q$ represents an irreducible fraction.
Suitable fractions for different fractional revivals are found by slicing the
circle depicted in \fref{fig_fractional}(a).
For a full revival the autocorrelation function for $p/q=0$ needs to be studied, 
whereas from  $p/q=1/2$ mirror revivals are identified.
The next higher fractional revivals with a triangular shape have the time shifts
$p/q=0$, $p/q=1/3$, $p/q=2/3$ and $p/q=1/2$, $p/q=1/6$, $p/q=5/6$.
The absolute value of the autocorrelation for these times is shown in
\fref{fig_fractional}(b) and indicates the times when revivals occur. 
A full revival takes place if the absolute value for $p/q=0$ becomes unity again, which is
the case for $t=T_\mathrm{rev}$.
The revival for $t=1/2\,T_\mathrm{rev}$ has a minimum for $p/q=0$ and a maximum for
$p/q=1/2$, explaining the discrepancy between the classical centre-of-mass and the quantum
mechanical result.
Fractional revivals with two peaks are visible at times $t=1/4\,T_\mathrm{rev}$ and
$t=3/4\,T_\mathrm{rev}$.
In this case the absolute value of the autocorrelation function has to reach almost $\sqrt{1/2}$.
Fractional revivals with a triangular shape can occur in two different ways.
For the first one a maximum is located at the opposite side of the initial starting point.
In this case the autocorrelation function for $p/q=1/2$, $p/q=1/6$, $p/q=5/6$ is $\sqrt{1/3}$
as shown in the numerical data.
As second possibility for a triangular revival the same result occurs for
$p/q=0$, $p/q=1/3$, $p/q=2/3$.
Also the next higher revivals can be seen in the numerical data but are not
presented here to keep clarity. 

\subsection{Effect of impurities}

For graphene based devices, a good understanding of the effect of impurities
on the electronic motion is required. Suspended graphene has a high mobility leading
to mean free path lengths of about $2\,\mu\mathrm{m}$~\cite{bolotin:096802}.
However, the mobility is reduced by the formation of intrinsic ripples which stabilise the two-dimensional 
system~\cite{Fasolino07}.
The lattice deformations involved can be added to the Dirac Hamiltonian by a position dependent gauge
field~\cite{Lammert00, Cortijo07}, which can be handled  by the propagation algorithm
described in section \ref{wppropagation}.
Additionally the curvature leads to a spin orbit interaction~\cite{huertas-hernando:155426}.
Here, we study the effect of a scalar potential which is commonly used to model impurities~\cite{Rycerz07}.
We use a simple model with a Gaussian distributed random potential and analyse the
propagation of a cyclotron wave-packet through this perturbed environment.
The potential was generated by the convolution of a random potential $U_n$ 
at the grid points $\myvec r_n$ of a square lattice:
%
%
%%%%%%%%%%
\begin{equation}
   U_\mathrm{imp}(\myvec r)=
	\sum_n U_n \exp \left ( 
	- \frac{|\myvec r - \myvec r_n |^2}{2 \xi^2}
	\right )
\mathrm{.}
\end{equation}
%%%%%%%%%%
%
%
The potential has a vanishing average expectation value
$\langle U_\mathrm{imp}(\myvec r) \rangle = 0$,
a variance of
$\langle U_\mathrm{imp}(\myvec r)^2 \rangle = U_0^2$,
and is correlated by
$\langle U_\mathrm{imp}(\myvec r) U_\mathrm{imp}(\myvec r') \rangle = U_0^2 \exp \left ( - |\myvec r - \myvec r'|^2/(2 \xi^2) \right )$
with the correlation length $\xi$.
A similar potential has already been used in conductance calculations
for graphene on a honeycomb lattice~\cite{Rycerz07}.
We focus on cases where the correlation length $\xi$ is
longer than the grid spacing leading to a potential which is a smooth function
with respect to the graphene unit cell.
%
% Fidelity
%%%%%%%%%
\begin{figure}[t]
 \centering \includegraphics[width=0.75\columnwidth]{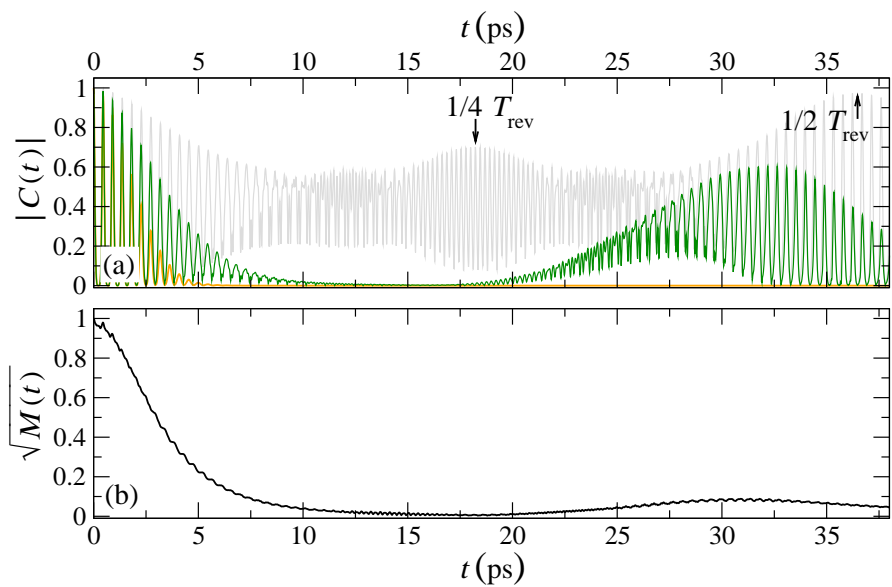}
   \caption{(a) Comparison between the autocorrelation function of two different impurity configurations
        (orange and green line, $\xi = 42\,\mathrm{nm}$, $U_0=5\,\mathrm{meV}$) and the
	autocorrelation function of the clean system (grey line).
	The numerical data shows that the potential strongly suppresses $|C(t)|$ 
	of a cyclotron wave-packet ($\bar n = 40$, $\sigma=1$)
	after a few orbits in a magnetic field of $B=10\,\mathrm{T}$.
        Revivals in the green line are not due to the phase relations between the addends
	of equation \eref{approxC}  but stem from a global motion of the wave-packet which returns by
	chance to the initial position.%
	(b) The fidelity $M(t)$ averaged over 10 different impurity configurations for the same initial cyclotron
	wave-packet ($\bar n = 40$, $\sigma=1$, $B=10\,\mathrm{T}$)
	feature a similar decay.
	\label{fig_fidelity}
	}
\end{figure}
%%%%%%%%%%%%%%%
%
%
In the following, we study several impurity configurations and calculate the propagation of a wave-packet
which has the same initial shape as a Landau level wave-packet of the clean system.
As first observable we investigate the autocorrelation function.
All the calculated results have in common that the autocorrelation function vanishes much faster
compared to the clean system.
In \fref{fig_fidelity}(a), the grey line shows the autocorrelation function of the clean system with prominent revivals
whereas the autocorrelation function in the presence of an impurity potential decays after a few picoseconds.
This phenomenon happens for many different setups since the wave-packet simply drifts away.
Some impurity configurations lead to a wave-packet motion, where the wave-packet returns to its initial position
after some time. The return results in an increased autocorrelation function but differs from the revivals occurring in the clean systems, which reflect the initial occupation of the Landau levels.

As a measure for the deviation of a disordered system from the clean system we calculate the fidelity
%
%
%%%%%%%%%%
\begin{equation}
   M(t) = | \langle \psi(0) | \e^{\ci \hat \mathcal H' t/\hbar} \e^{-\ci \hat \mathcal H t/\hbar} | \psi(0) \rangle |^2
   \mathrm{.}
\end{equation}
%%%%%%%%%%
%
%
We track the overlap of the wave-packet propagated by the unperturbed Hamiltonian $\hat \mathcal H$
and the perturbed Hamiltonian $\hat \mathcal H' = \hat \mathcal  H + U_\mathrm{imp}$.
For the free electron gas, expressions are known which describe the fidelity decay in
disordered systems for differently correlated impurity potentials~\cite{PhysRevE.67.056217}.
We obtain similar numerical results, shown in \fref{fig_fidelity}(b).
For short times the fidelity is approximately unity and then decays exponentially.
After longer times, the fidelity does not approach zero since there is a non-vanishing
probability to return back to the initial position.
However we cannot infer from the autocorrelation function whether wave-packet revivals occur or not,
since the exponential fidelity decay removes all relevant information from the calculated data.

%
% CM Trajectories with impurity
%%%%%%%%%
\begin{figure}[t]
 \centering \includegraphics[width=0.75\columnwidth]{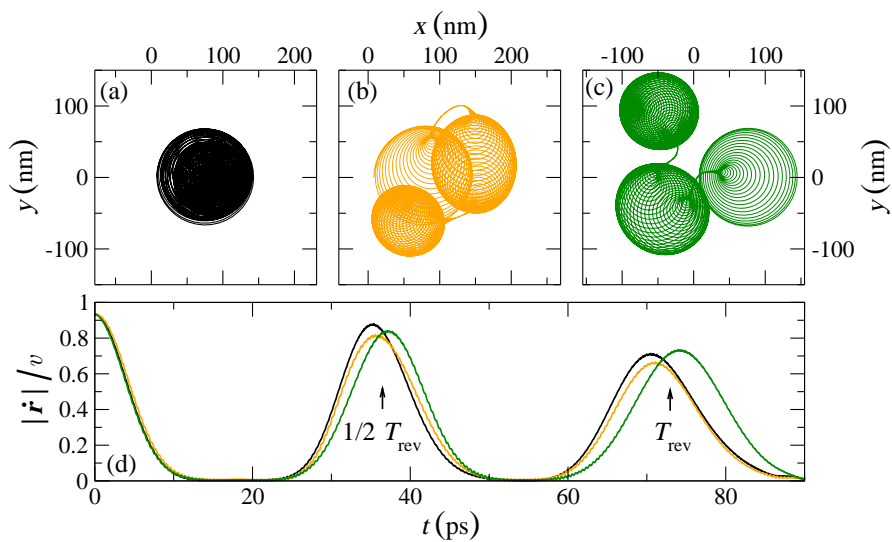}
   \caption{(a-c) Centre-of-mass motion of a cyclotron wave-packet
	($\bar n = 40$, $\sigma=1$, $B=10\,\mathrm{T}$) for different impurity configurations
        ($\xi = 42\,\mathrm{nm}$, $U_0=5\,\mathrm{meV}$).
	The time-evolution shows the collapse of the localised wave-packet by a shrinking spiral.
	The evenly distributed circular wave-packet shows much smother movement 
	leading to a connecting line between two revivals.
	(d) Velocity of the centre-of-mass for the depicted wave-packet	trajectories from above.
	The velocity approaches the ``speed of light'' of graphene for each revival.
	Revivals appear at the same times as in the clean system.
	Although the impurity potential reduces the velocity of the wave-packet for $1/2 T_\mathrm{rev}$
	and $T_\mathrm{rev}$ the effect of revivals is still very pronounced in the time-evolution
	of the centre-of-mass velocity.
	\label{fig_imp_CM}
	}
\end{figure}
%%%%%%%%%%%%%%%
%
%
Consequently we have to find other observables which signify revivals
even if an impurity potential is present.
One particularly suitable observable is the centre-of-mass motion of the wave-packet.
In \fref{fig_imp_CM}(a-c) the centre-of-mass motion of a Dirac wave-packet is shown
for different impurity potentials.
In \fref{fig_imp_CM}(a) the wave-packet does not drift away and thus the revivals
lie on top of each other.
This leads to a revival in the autocorrelation function which has nothing in common with the
wave-packet revivals of the clean system since the revival time depends crucially on the
exact choice of the impurity potential.
%
%
% Snapshots with impurity
%%%%%%%%%
\begin{figure}[t]
 \centering \includegraphics[width=0.75\columnwidth]{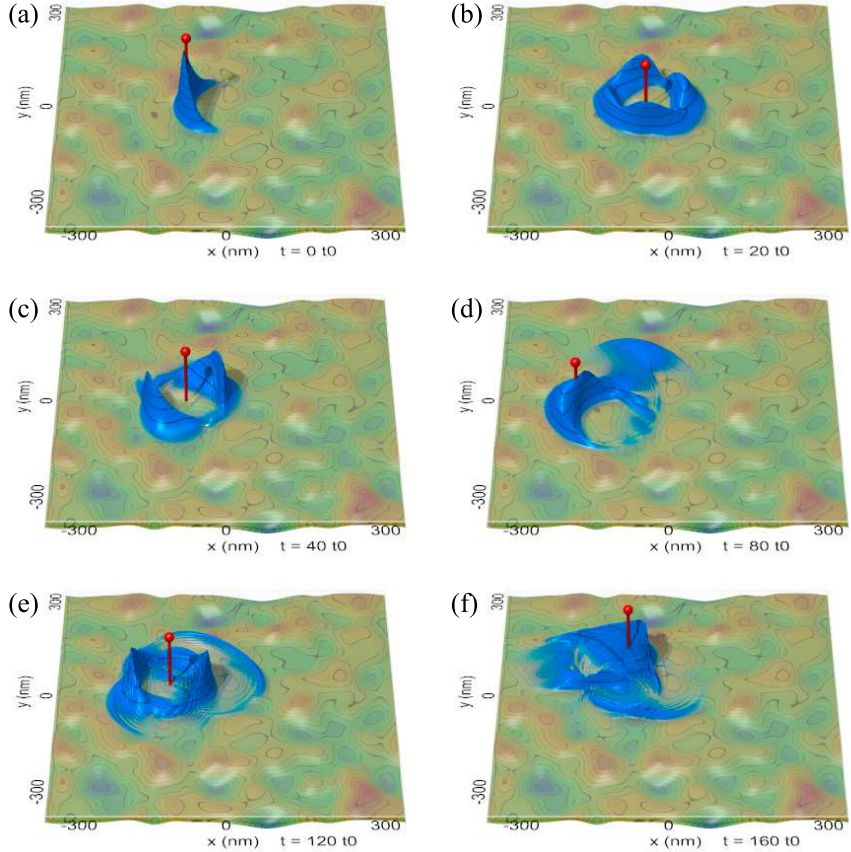}
   \caption{Revivals and fractional revivals of a cyclotron wave-packet in an impurity potential
        ($\xi = 42\,\mathrm{nm}$, $U_0=5\,\mathrm{meV}$).
	The red pin marks the quantum mechanical centre-of-mass.	
	(a) The initial wave-packet with a Gaussian eigenstate contribution 
	$\bar n = 40$, $\sigma = 1$.
	$B=10\,\mathrm{T}$ leads to a Poincar\'e cyclotron time of $t0=T_\mathrm{cl}$.
	During the time evolution the centre-of-mass collapses to the centre of a circular
	wavefunction featuring pronounced revivals. In
	(b) a quarter revival occurs at $t= 1/2 \bar n\,T_\mathrm{cl}$ and in 
	(c) a half revival at $t= \bar n\,T_\mathrm{cl}$.
	After a certain time a localised wave-packet emerges again and 
	(d) shows a mirror revival for $t= 2 \bar n\,T_\mathrm{cl}$.
	In this case some part of the wavefunction is detached from the
	localised cyclotron wave-packet.
	(e) The centre-of-mass collapses again and forms fractional revivals.
	(f) The full revival where the semiclassical and the quantum mechanical centre-of-mass
        coincide occurs at $t= 4 \bar n\,T_\mathrm{cl}$.
	After this time the wavefunction shows pronounced branches.
	However, the probability distribution is clumped around the centre-of-mass showing the
	characteristics of a revival in the velocity expectation value.
	\label{fig_impurity_snapshots}
	}
\end{figure}
%%%%%%%%%%%%%%%
%
%
However the majority of the disorder configurations lead to a drift of the wave-packet as shown in
\fref{fig_imp_CM}(b,c).
In all cases the wave-packet moves at the beginning on a shrinking spiral since the
centre-of-mass is located at the position with the highest probability density
(\fref{fig_impurity_snapshots}(a)).
In this state the centre of the spiral already moves. 
After a few picoseconds the centre-of-mass collapses to the centre of a rotating circular
wave-packet as shown in \fref{fig_impurity_snapshots}(b,c).
In the following the wave-packet is moving without oscillations
on a smooth line because the density is equally distributed along a circular orbit.
As in the clean system, the first revival occurs in the form of a mirror revival at time
$1/2\,T_\mathrm{rev}$.
Then the centre-of-mass collapses again and produces a full revival at $T_\mathrm{rev}$.
A way to remove the structure of the impurity potential is extracting the corresponding velocity of the wave-packet  
out of the centre-of-mass motion
%
%
%%%%%%%%%%
\begin{equation}
  \myvec v = \frac{\partial}{\partial t}
    \langle \psi(t) | \hat \myvec r | \psi(t) \rangle
     \mathrm{,}
\end{equation}
%%%%%%%%%%
%
%
which is shown in \fref{fig_imp_CM}(d).
The initial wave-packet moves on the circular orbit with a velocity slightly smaller than the
``speed of light'' of graphene.
When the centre-of-mass collapses this velocity drops down to the drift velocity of the wave-packet
which is defined by the potential gradient and at least two orders of magnitude smaller.
The revivals occur at the same time for all the impurity configurations.
The substructure of the revivals cannot be seen in the centre-of-mass motion of the wave-packet.
However the full wavefunction shows fractional revivals (\fref{fig_impurity_snapshots}).

In reference~\cite{rusin-2008} the centre-of-mass propagation was
used to establish a dipole moment.
In our case this dipole moment is maximal if the wave-packet revival takes place and
almost zero if the wave-packet is collapsed similar to the velocity of the wave-packet in
\fref{fig_imp_CM}(d).
If such a dipole moment is experimentally measurable, the wave-packet revivals can be observed
even if disorder displaces the wave-packet far away from its initial position.

\section{Conclusion}

In this work we presented a new algorithm for the propagation of wave-packets on a sheet of graphene 
which allows us to calculate the time-evolution in arbitrary shaped potentials and
inhomogeneous magnetic fields.
We studied time-dependent effects like the {\it zitterbewegung},
which is very pronounced in graphene.
We focussed on wave-packet revivals, which occur in graphene due to the special structure of the Landau levels.
We gave a detailed explanation of the centre-of-mass motion and the autocorrelation function.
The applied semiclassical description quantitatively revealed the collapse of the centre of
mass as well as the revivals.
From the autocorrelation function we extracted several different timescales which all signify various
revivals.
A detailed analysis of the autocorrelation function for different Poincar\'e times confirmed
fractional revivals which have also been seen in our numerical calculations which supported the results
throughout the whole work.
Finally we analysed the effect of impurities on the wave-packet revivals by a Gaussian
correlated model potential.
Our calculations showed that revivals in the autocorrelation are strongly suppressed
due to the fidelity decay.
In contrast to that the centre-of-mass revivals are still present
if the correlation length of the impurity potential is long enough with respect
to the average cyclotron radius.
The here demonstrated long-time accuracy of the propagation algorithm opens
the window towards realistic device simulations on the micrometre range, also including
time-dependent external fields.

\ack
We thank
C.\ Petitjean,
C.\ Kreisbeck,
J.\ Wurm,
M.\ Wimmer,
M.\ Hartung,
E.\ J.\ Heller,
C.\ Koch,
K.\ Richter,
J.\ Schliemann, and
W.\ P.\ Schleich
for helpful discussions.
This work is supported by the Emmy-Noether programme of the DFG (KR 2889-2/1).

\appendix

\section{Classical Propagation}\label{appendixA}

In the following we study the classical time-evolution of a wave-packet which is located
around one of the Dirac points and obtain the semiclassical quantisation.
In order to be localised in position and momentum space, the initial
wave-packet $\psi(\myvec r, t)$ has to have a small extent in both coordinates.
The expectation value of the position operator and the kinetic momentum are defined by
%
%
%%%%%%%%%%
\begin{equation}
  \myvec r(t) =
   \langle \psi(t) | \hat{\myvec r} | \psi(t) \rangle 
   \mathrm{,\ \ \ \ }
  \myvec k(t) =
   \langle \psi(t) | \hat{\myvec k} | \psi(t) \rangle 
   \mathrm{.}
\end{equation}
%%%%%%%%%%
%
%
Niu {\it et. al.} have developed a framework to study the time-evolution of those
observables for arbitrary Hamilton operators~\cite{PhysRevB.59.14915} which we will apply
in the following.
As a first step we analyse the free solutions of the Hamiltonian and classify the 
resulting bands by the index $\lambda$.
For the Dirac Hamiltonian of graphene we have electron-like ($\lambda = +1$)
and hole-like states ($\lambda = -1$) with the corresponding energy dispersion
$\mathcal E_\lambda = \lambda~v~\hbar |\myvec k|$.
Then it is possible to construct a wave-packet which is restricted to one band.
The time-evolution of the momentum and the centre of this wave-packet are defined
by two coupled equations of motion
%
%
%%%%%%%%%%
\begin{eqnarray}
\dot{ \myvec r} &=& \frac{\partial {\mathcal E}_\lambda}{\hbar \partial k} - \dot{\myvec k} 
 \times {\bf \Omega_\lambda}(\myvec k)
 \mathrm{,}\label{EqofMr}\\
\hbar \dot{\myvec k} &=& -e \myvec E(\myvec r) - e \dot{\myvec r} \times \myvec B
 \mathrm{,}\label{EqofMk}
\end{eqnarray}
%%%%%%%%%%
%
%
where  $\myvec B = B \myvec e_z$ stands for an isotropic magnetic field perpendicular
to the plane and $\myvec E(\myvec r)$ is a position dependent electric potential.
The Berry curvature ${\bf \Omega}_\lambda(\myvec k)$ enters the equations of motion in momentum space
but is zero in the case of a gap-less perfect sheet of graphene.
As a result of the linear dispersion, relation \eref{EqofMr} reduces
to $\dot{ \myvec r} = \lambda v \myvec k / |\myvec k|$
and therefore describes a centre-of-mass  which always travels with the speed
$v$ parallel or antiparallel to the wave vector $\myvec k$.
For a vanishing electric field the two coupled equations of motion can be solved.
The momentum of the wave-packet is controlled by the differential equation
%
%
%%%%%%%%%%
\begin{equation}
 \hbar \dot{\myvec k} = - e v \frac{\myvec k}{|\myvec k|} \times \myvec B(\myvec r)\mathrm{,}
\end{equation}
%%%%%%%%%%
%
%
given by \eref{EqofMk} which is solved by
%
%
%%%%%%%%%%
\begin{equation}
  \myvec k = k_0   \left (
    \begin{array}{c}
      \cos(\omega_c t + \varphi) \\
      \sin(\omega_c t + \varphi)
     \end{array}
  \right )
  \mathrm{,}
\end{equation}
%%%%%%%%%%
%
%
where the initial momentum of the wave-packet is given by $k_0$ 
in direction $\varphi$. The cyclotron frequency depends on the momentum and is
given by
%
%
%%%%%%%%%%
\begin{equation}
  \omega_c = \frac{e v B}{\hbar k_0} \mathrm{.}
\end{equation}
%%%%%%%%%%
%
%
The resulting real space propagation \eref{EqofMr} is given by
%
%
%%%%%%%%%%
\begin{equation}
  \myvec r =  \myvec r_0 + \lambda\;l_c   \left (
    \begin{array}{c}
      \sin(\omega_c t + \varphi) \\
      -\cos(\omega_c t + \varphi)
     \end{array}
  \right ) \mathrm{.}
\end{equation}
%%%%%%%%%%
%
%
Thus electron-like and hole-like states propagate in opposite directions on orbits
with the radius
%
%
%%%%%%%%%%
\begin{equation}
  l_c = \frac{\hbar k_0}{e B} \mathrm{.}
\end{equation}
%%%%%%%%%%
%
%
In order to obtain the quantisation of the Landau levels from the closed
cyclotron orbits we have to study the phase changes along the trajectories.
The main contribution is given by the classical action known from the electron case, while 
an additional contribution comes from Berry's phase.
Although the Berry curvature is vanishing, Berry's phase~\cite{Berry84} is still present
and defined by the vector potential
%
%
%%%%%%%%%%
\begin{equation}
  \mathcal{\vec A}_\lambda(\myvec k) = -\ci
    \langle u_\lambda (\myvec k) | \nabla_k  u_\lambda (\myvec k ) \rangle
  \mathrm{,}
\end{equation}
%%%%%%%%%%
%
%
and the spinor parts $u_\lambda (\myvec k)$ of the free solutions of the Hamiltonian.
The free solutions are given by
%
%
%%%%%%%%%%
\begin{equation}
  u_\pm (\myvec k) = \frac{1}{2}\sqrt{2}
   \left (
    \begin{array}{c}
	1\\
	\e^{\pm \ci \varphi}
     \end{array}
  \right )
  \mathrm{,}
\end{equation}
%%%%%%%%%%
%
%
where $\varphi$ is the direction of the vector $\myvec k$.
Hence we get a vector potential of
%
%
%%%%%%%%%%
\begin{equation}
  \mathcal{\vec A}_\pm(\myvec k) = \pm \frac{1}{2 k^2}
  \left (
    \begin{array}{c}
	k_y\\
	-k_x
     \end{array}
  \right )
  \mathrm{,}
\end{equation}
%%%%%%%%%%
%
%
which is used to calculate the additional phase-change
%
%
%%%%%%%%%%
\begin{equation}
  \Gamma_\lambda (\omega) = \oint_\omega \mathcal{\vec A}_\lambda(\myvec k) d \myvec k
\end{equation}
%%%%%%%%%%
%
%
for an arbitrary adiabatic transition on the path $\omega(t)$ in momentum space.
The result is Berry's phase
%
%
%%%%%%%%%%
\begin{equation}
  \Gamma_\pm (\omega) = \pm \pi
\end{equation}
%%%%%%%%%%
%
%
for a single closed orbit on graphene~\cite{Kim2005}.
The adiabatic transition in momentum space is automatically fulfilled in a magnetic field,
since all trajectories are cyclotron orbits and thus sufficiently smooth.
The classical phase change and Berry's phase have to be added up to build a new
EBK quantisation condition which reads~\cite{PhysRevLett.75.1348}
%
%
%%%%%%%%%%
\begin{equation}
  \frac{1}{2} \myvec e_z \oint_\omega \myvec k \times d \myvec k  = 
        2 \pi \frac{e B}{\hbar}
        \left ( n' + \frac{\nu}{4} - \frac{\Gamma_\pm (\omega)}{2 \pi} \right)
\end{equation}
%%%%%%%%%%
%
%
where $\nu$ is the Maslov index depending on the caustics traversed on one cyclotron orbit
(here $\nu=2$). Plugging the formula for the cyclotron motion in the quantisation
condition the resulting Landau levels are given by
%
%
%%%%%%%%%%
\begin{equation}
 \pi k_0^2 = 
    2 \pi \frac{e B}{\hbar} \left ( n' + \frac{1}{2} \mp \frac{1}{2} \right)\mathrm{.}
\end{equation}
%%%%%%%%%%
%
%
If we define the Landau level index $n$ differently for hole-like and electron-like states
the momentum quantisation yields
%
%
%%%%%%%%%%
\begin{equation}
k_0 = \mathrm{sign}(n) \sqrt{2 \frac{e B}{\hbar} |n|}
\mathrm{.}
\label{quantization_condition}
\end{equation}
%%%%%%%%%%
%
%
All the other quantised parameters follow from the relations calculated before
%
%
%%%%%%%%%%
\begin{equation}
\mathcal E_n = v~\mathrm{sign}(n) \sqrt{2 e B \hbar |n|}
\mathrm{,}
\label{quantized_E}
\end{equation}
%%%%%%%%%%
%
%

%
%
%%%%%%%%%%
\begin{equation}
l_n = \mathrm{sign}(n) \sqrt{2 \frac{\hbar}{e B} |n|}
\mathrm{,}
\label{quantized_l}
\end{equation}
%%%%%%%%%%
%
%

%
%
%%%%%%%%%%
\begin{equation}
  \mathcal \omega_n = v~\mathrm{sign}(n) \sqrt{\frac{e B}{2 \hbar |n|} }
\mathrm{\ \ if \ \ }
  n \neq 0
\mathrm{.}
\label{quantized_w}
\end{equation}
%%%%%%%%%%
%
%

\section*{References}
%\bibliographystyle{iop}
%\bibliographystyle{unsrt}
%\bibliographystyle{vancouver}
%\bibliography{refs.bib}
\providecommand{\newblock}{}

\end{document}